\documentclass[12pt]{article}
\usepackage{a4wide}
\usepackage[dvips]{graphicx,color}
\usepackage{amsmath,latexsym,amssymb,amsfonts}
\usepackage{bm}
\usepackage{hyperref}

\begin{document}
	{\renewcommand{\thefootnote}{\fnsymbol{footnote}}

\begin{center}
	{\LARGE Can a false vacuum bubble remove the singularity inside a black hole?}\\
	\vspace{1.5em}
	Suddhasattwa Brahma$^{1,2}$\footnote{e-mail address: {\tt suddhasattwa.brahma@gmail.com}}  
	and Dong-han Yeom$^{3,4}$\footnote{e-mail address: {\tt innocent.yeom@gmail.com}}
	\\
	\vspace{0.5em}
	{\small $^1$Asia Pacific Center for Theoretical Physics, Pohang 37673, Republic of Korea}\\
	{\small $^2$Department of Physics, McGill University, Montr\'{e}al, QC H3A 2T8, Canada}\\
	{\small $^3$Department of Physics Education, Pusan National University, Busan 46241, Republic of Korea}\\
	{\small $^4$Research Center for Dielectric and Advanced Matter Physics, Pusan National University, Busan 46241, Republic of Korea}\\
	\vspace{1.5em}
\end{center}
}

\setcounter{footnote}{0}

\begin{abstract}
We investigate a \textit{regular} black hole model with a de Sitter-like core at its center. This type of a black hole model with a false vacuum core was introduced with the hope of singularity-resolution and is a common feature shared by many regular black holes. In this paper, we examine this claim of a singularity-free black hole by employing the thin-shell formalism, and exploring its dynamics, within the Vaidya approximation. We find that during gravitational collapse, the shell necessarily moves along a space-like direction. More interestingly, during the evaporation phase, the shell and the outer apparent horizon approach each other but, unless the evaporation takes place very rapidly, the approaching tendency is too slow to avoid singularity-formation. This shows that albeit a false vacuum core may remove the singularity along the ingoing null direction, there still exists a singularity along the outgoing null direction, unless the evaporation is very strong.
\end{abstract}

%
%

\section{Introduction}
One of the most challenging problems in fundamental physics remains the appearance of classical singularities within the cores of black holes \cite{Hawking:1969sw}. Singularity theorems in general relativity (GR) \cite{Hawking:1973uf} generically leads to such regions where the classical notions of spacetime stop being meaningful. The general consensus is that, near Planckian curvatures, a theory of quantum gravity should supersede classical GR thereby resolving the singularity. For instance, in a vast array of quantum cosmological models \cite{Ashtekar:2011ni}, one finds that the curvature invariants reach a maximum, without blowing  up to infinity, due to the appearance of quantum gravity corrections. In the context of black holes, one is then led to ask the natural question: which one of the assumptions of the singularity theorems \cite{Chen:2014jwq} --  (1) the existence of an apparent horizon, (2) global hyperbolicity, or (3) the null energy condition -- is violated so as to bypass them? However, in order to understand exactly how a specific quantum gravity approach violates one of these three assumptions \cite{AyonBeato:1999rg}, we need to further assume that the quantum theory allows for an \textit{effective spacetime} approach. This can be understood as follows: It is entirely possible that singularity-resolution is a completely `quantum' process which requires solving the Wheeler de-Witt equation \cite{DeWitt:1967yk}, or some regularized version of it \cite{Ashtekar:2011ni}, for the wave function and thereby evaluate physical observables from it. However, derivation of effective spacetimes is the further assumption that one can choose suitable semiclassical states within the quantum theory and calculate the expectation values of relevant geometric operators such that one can regard the an effectively classical geometry near the singularity, with quantum gravity effects appearing as corrections to the classical field equations. If such an effective geometry removes the singularity, one is left with a so-called a \textit{regular} black hole model \cite{Bardeen}. Although such a nonsingular effective description is not guaranteed to appear even if a quantum gravity theory exists, in the absence of a full description of any such fundamental quantum gravity approach, such regular black hole models are useful first examples to examine if classical paradoxes can get resolved in such cases. 

The information loss paradox is one such problem which arises as a consequence of the classical singularity mentioned above. In order to understand this and other such problems, it is useful to investigate the \textit{dynamical} causal structures of regular black hole models by studying the \textit{effective} quantum gravitational completion of the evaporating spacetime \cite{Frolov:1988vj,Hayward:2005gi}. For such black holes, the first condition -- the existence of an apparent horizon -- is necessary in order to construct the formation of an asymptotically Schwarzschild spacetime, therefore leaving us no way to avoid this. A popular next choice has been to violate the null energy condition \cite{Bronnikov:2005gm}, but this typically comes with its own plethora of problems \cite{Cline:2003gs}. Interestingly, there do exist regular black hole models which satisfy the null energy condition \cite{Bardeen,Frolov:1988vj,Hayward:2005gi}. The essential point of these models is the violation of global hyperbolicity \cite{Borde:1996df}, perhaps the most conservative of the three options. This implies that the center of the black hole is time-like and there should exist a Cauchy horizon rendering the center to be regular. Nevertheless, even on choosing this option, the existence of the Cauchy horizon leads to the problem of instability and mass inflation of the inner horizon \cite{Poisson:1990eh}.

Even apart from the inner horizon issue, there still remains a problem. If the black hole spacetime is dynamical, the violation of global hyperbolicity becomes subtle. In principle, both the formation and evaporation processes can be well controlled by the given initial data for a classical black hole. However, for regular black holes with a Cauchy horizon, one wonders if the regularity of the solution comes at the cost of having a truly dynamical description? In other words, are such regular black holes only toy models which cannot to be extended to modeling realistic black holes? If it is not to be the case, what is then `effectively' violated among the three assumptions of the singularity theorem? A naive answer is that if the black hole spacetime is dynamical, due to the Hawking radiation, the null energy condition is violated. Hence, we are led back to this option in order not to have formation of a singularity even in the case of dynamical (regular) black holes. Eventually, one may obtain a closed apparent horizon with an entirely regular and time-like center \cite{Hayward:2005gi}. This is a general belief exhibited in the vast literature on regular black holes.

In this paper, we critically revisit this interpretation about regular black holes. In order to see the internal structures in detail, we need a way to dynamically probe these structures. However, if the model is an effective model and only the static solution is known, there is no unambiguous and justified way to generalize such internal structures to the dynamical case. In order to avoid this difficulty, we assume that there is an evaporating black hole, where the outside is Schwarzschild and the inside is a dS-like core. Note that the dS-like core at the center is a very generic properties of a wide variety of regular black holes in the literature \cite{Bardeen,Frolov:1988vj, Hayward:2005gi}. Once we assume the dS core, in principle we can introduce the field theoretical model \cite{Hansen:2009kn} and estimate the dynamics of the transient region by numerical methods  \cite{Hwang:2012nn} or through other similar approximations.

Building on this hypothesis, we may go one step further by assuming that (1) the outside is Schwarzschild and inside is dS, (2) the intermediate region is described by energy-momentum tensors which satisfy the \textit{dynamical} Einstein's equation (as opposed to just having a metric ansatz), and (3) the space-time is globally hyperbolic (at least, up to the numerically calculated domain). For simplicity, and to keep the calculations tractable, we assume that the energy-momentum tensor will be approximated by thin-shells. This allows us to investigate quite generic, field-theoretically justifiable regular black hole dynamics. As an unexpected conclusion, we correct some naive expectations of regular black hole models and point out overlooked issues in the typical causal structures of an evaporating regular black hole\footnote{Other problems with regular black hole models have also been pointed out in the literature; see, e.g. \cite{Carballo-Rubio:2018pmi}.}. This lets us better formulate the following question: What is the genuine condition to obtain closed apparent horizons?

This paper is organized as follows. In Sec.~\ref{sec:mod}, we describe mathematical details of the model. In Sec.~\ref{sec:dyn}, we obtain and analyze numerical solutions. In Sec.~\ref{sec:qgc}, we comment on the limitation of the thin-shell description and possible genuine quantum gravitational modifications. Finally, in Sec.~\ref{sec:con}, we summarize and discussion about possible future research directions.

\section{\label{sec:mod}The model}
In this section, we describe the regular black hole system modeled by a thin-shell, where inside the shell is de Sitter (dS) and outside the shell is Schwarzschild. In order to introduce radiation and make the system dynamical, we choose the Vaidya metric outside the shell where the mass parameter can vary with time. The reason for choosing such a model is so that we recover an asymptotically flat spacetime for large areal radius $r$, whereas the dS core ensures the flatness condition at the center. On the other hand, choosing the thin-shell formalism ensures that the model retains its proximity of interpreting the results in terms of field-theoretic models of black hole evaporation. Indeed, a major improvement of this type of a setup, when compared with similar dynamical regular black hole models, lies in its goal of connecting it to realistic evaporation scenarios.

\subsection{Metric ansatz}
As mentioned, in order to investigate dynamical properties of regular black holes,  we consider a thin-shell, the region exterior to which is given by
\begin{eqnarray}\label{eq:metout}
ds_{+}^{2} = - \left(1 - \frac{2 m(V)}{r} \right) dV^{2} + 2dVdr + r^{2} d\Omega^{2}\,,
\end{eqnarray}
whereas the interior is described as
\begin{eqnarray}
ds_{-}^{2} = - \left(1 - \frac{r^{2}}{\ell^{2}} \right) dv^{2} + 2dvdr + r^{2} d\Omega^{2}\,.
\end{eqnarray}
The outside region is described by a Vaidya spacetime whereas $V$ and $v$ denote incoming null coordinates for outside and inside of the shell, respectively. The mass depends on the time as parametrized by $m(V)$ and $\ell$ denotes the Hubble radius of the internal de Sitter space. We consider that the shell is located at $r = R(v)$, and from Israel's boundary condition, the two metrics $ds_{-}^{2}$ and $ds_{+}^{2}$ must be smoothly continuous at the shell. By introducing a new coordinate $z$,  \cite{Chen:2017pkl}
\begin{eqnarray}
	r = \begin{cases}
	R + \frac{z}{V'}\;\; (z>0)\,,\\
	R + z \;\; (z\leq 0)\,,
	\end{cases}
\end{eqnarray}
so that the position of the thin-shell in terms of this new coordinate is given at $z=0$. Here, prime denotes a derivative with respect to $v$. In terms of these variables, the spacetime metric looks as follows, which is continuous across the shell:
\begin{eqnarray}
ds^{2} &=& - \left[ \left( \left( 1 - \frac{2m}{R+z/V'} \right) V'^{2} + 2 z \frac{V''}{V'} - 2 V'R' \right)\Theta(z)  \nonumber \right.\\
&& \left.+ \left( \left( 1 - \frac{(R+z)^{2}}{\ell^{2}} \right) - 2 R' \right) \left( 1 - \Theta(z) \right) \right] dv^{2} \nonumber \\
&& + 2 dvdz + \left( R + \frac{z}{V'}\Theta(z) + z \left(1-\Theta(z)\right) \right)^{2} d\Omega^{2}\,.
\end{eqnarray}
Continuity of the induced metric on the thin-shell leads to the following equation
\begin{eqnarray}\label{eq:cont}
R' = \frac{(1 - R^{2}/\ell^{2}) - (1 - 2m/R) V'^{2}}{2 ( 1 - V') }\,.
\end{eqnarray}

\begin{figure}
\begin{center}
\includegraphics[scale=0.7]{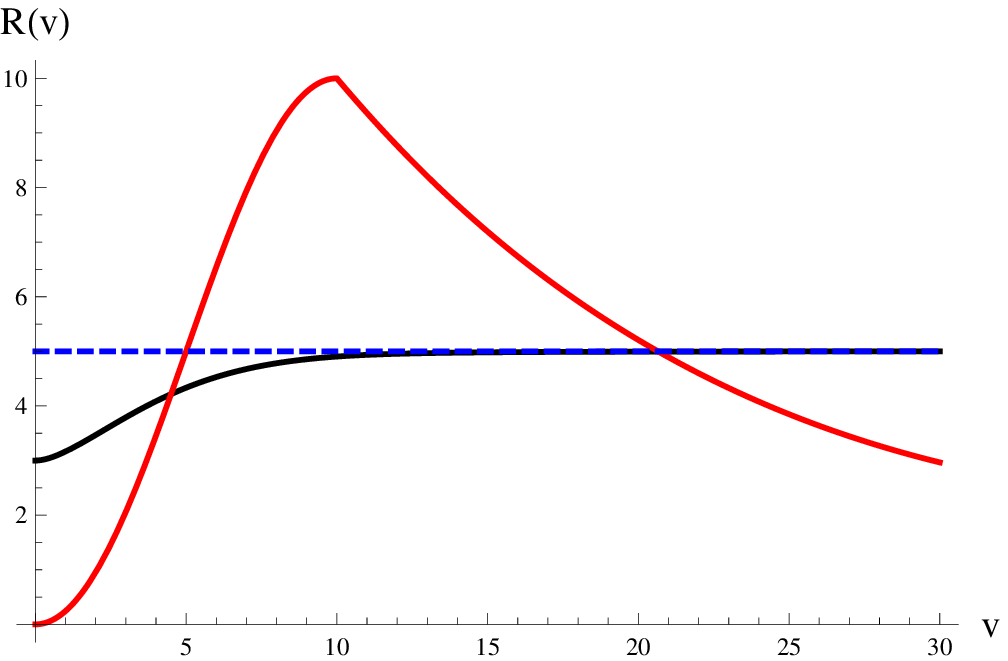}
\includegraphics[scale=0.7]{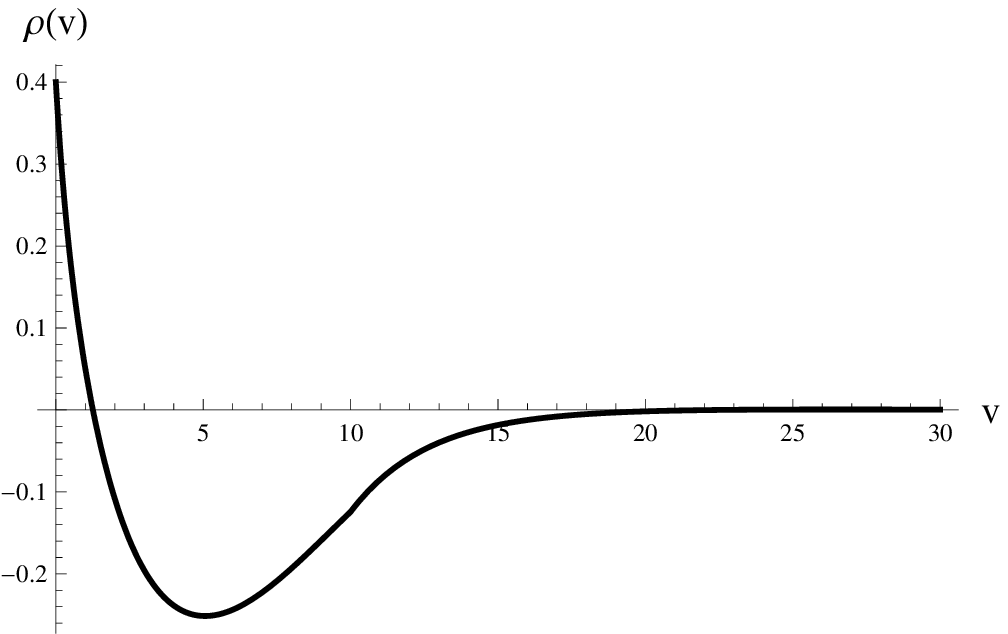}
\includegraphics[scale=0.7]{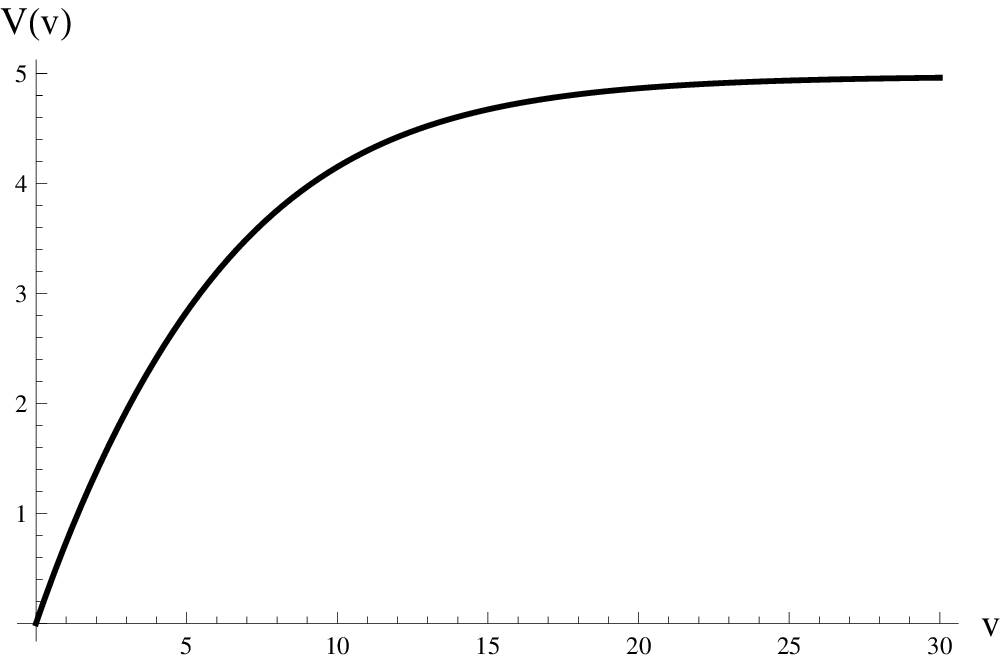}
\includegraphics[scale=0.7]{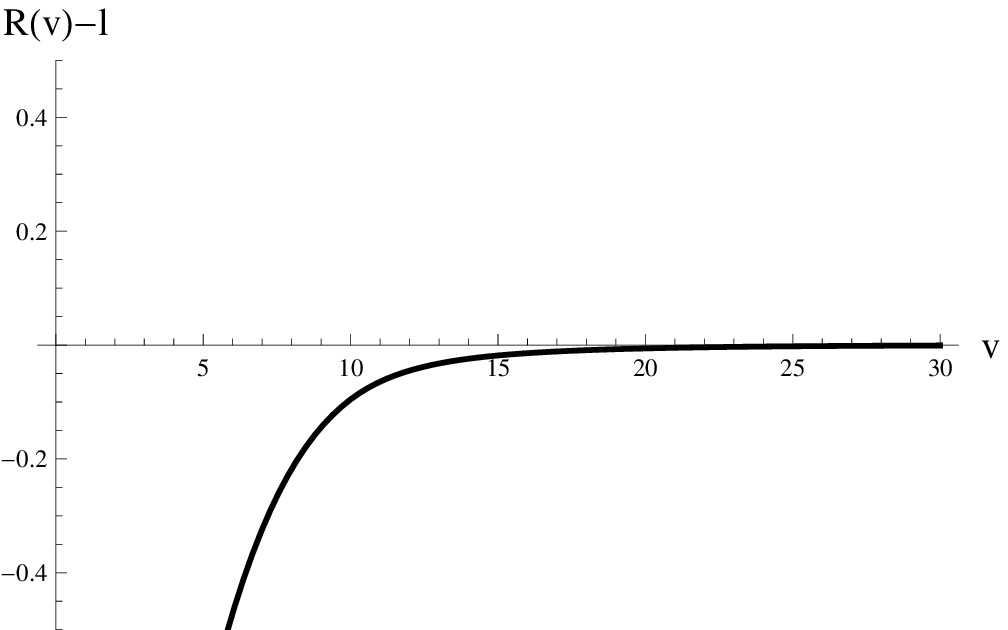}
\caption{\label{fig:type1}Numerical solution for $m_{0} = 5$, $\ell = 5$, $v_{0}=10$, $\alpha = 50$, $R(0) = 3$, $R'(0) = 0$, and $V(0) = 0$. Upper left: the shell trajectory (black thick curve), where the red curve is the (putative) outer apparent horizon ($R = 2m$) and the blue dashed curve is $R = \ell$; this outer apparent horizon is only meaningful if it is outside the shell. Upper right: $\rho(v)$, where the shell is time-like if $\rho > 0$ and space-like if $\rho < 0$. Lower left: $V(v)$ increases and approaches to a finite value as $v$ goes to infinity. Lower right: $R - \ell$ approaches to zero and hence the shell approaches to the horizon of the inside de Sitter space.}
\end{center}
\end{figure}

\subsection{Field Equations}
We adopt the formalism of Chen, Unruh, Wu and Yeom \cite{Chen:2017pkl}. By differentiating both sides of the Eq.~\eqref{eq:cont}, we get
\begin{eqnarray}
R'' &=& \frac{(1-V'^{2} (1-2m/R) - R^{2}/\ell^{2}) V''}{2 (1-V')^{2}} +\nonumber\\
&& \frac{- 2V' (1-2m/R) V'' - 2 R R'/\ell^{2} - V'^{2} (-2 m'/R + 2 m R'/R^{2})}{2 (1-V')}\,.
\end{eqnarray}
On the other hand, from the Einstein's equation, one can write the Einstein tensor as $G^{\mu\nu}_{\mathrm{bulk}} + \delta(z) G^{\mu\nu}_{\mathrm{shell}}$. Due to spherical symmetry, the non-zero components of the Einstein tensor are
\begin{eqnarray}
G^{vv}_{\mathrm{shell}} &=& \frac{2 (V'-1)}{RV'}\,,\\
G^{\theta\theta}_{\mathrm{shell}} &=& -\frac{2}{R\ell^{2}} - \left(\frac{RV'^{2} - mV'^{2} - RV' + V'' R^{2}}{R^{4} V'}\right)\,.
\end{eqnarray}
The first component is related to the tension of the shell and the second component is related to the tangential stress of the shell. If the Einstein equations are satisfied at the junction surface, then the continuity of not only the metric itself but also the first derivative of the metric is guaranteed.

By imposing the zero tangential stress condition, we can simplify all the equations of motion that describe dynamical behavior of the thin shell. On imposition of the zero tangential stress condition, i.e. assuming that the thin-shell is composed of dust, we impose
\begin{eqnarray}
G^{\theta\theta}_{\mathrm{shell}} = 0\,,
\end{eqnarray}
or, equivalently
\begin{eqnarray}
V'' = \frac{V'^{2}m - R V'^{2} + RV' - 2R^{3}V'/\ell^{2}}{R^{2}}\,.
\end{eqnarray}
Therefore, we end up with two second order differential equations for $R''$ and $V''$, which we go on to solve numerically. 

\section{\label{sec:dyn}Dynamics of the regular black hole core}
\subsection{Numerical solutions}
In order to find numerical solutions, we first need to postulate a specific form for $m(v)$ -- we consider that the mass increases from zero to $m_{0}$ during the time $0 \leq v \leq v_{0}$. After this time, the black hole begins to evaporate following the Hawking formula (or, at least, some Hawking-like formula). Our ansatz takes the form \cite{BMT}
\begin{eqnarray}
m'(v) &=& \frac{\pi m_{0}}{2 v_{0}} \sin \frac{\pi}{v_{0}}v \;\;\;\;\;\;\;\;\;\;\;\; 0 \leq v \leq v_{0}\,,\\
&=& - \frac{\alpha}{m^{2}} V' \;\;\;\;\;\;\;\;\;\;\;\;\;\;\;\;\;\;\; v_{0} \leq v\,,
\end{eqnarray}
where $\alpha$ is a constant that is proportional to the number of fields that contribute to Hawking radiation and $m(0) = 0$. Note that although $m'$ is discontinuous but there is no physical or mathematical problem for further investigations. The first part of this ansatz is just our mathematical choice, but the qualitative details will not depend on the form of $m'$ for $v \leq v_{0}$. In any case, it is easy to further refine the ansatz so as to make $m'$ continuous.

We need four more free parameters $R(0)$, $R'(0)$, $V(0)$, and $V'(0)$, where $V(0)$ is a parameter we are free to choose. One of the other three parameters will be fixed by satisfying Eq.~(\ref{eq:cont}). For consistency and convenience, we choose $0 \leq R(0) \leq \ell$ and, in the process, ensure that there is no horizon at the beginning of the collapse.

\begin{figure}
\begin{center}
\includegraphics[scale=0.7]{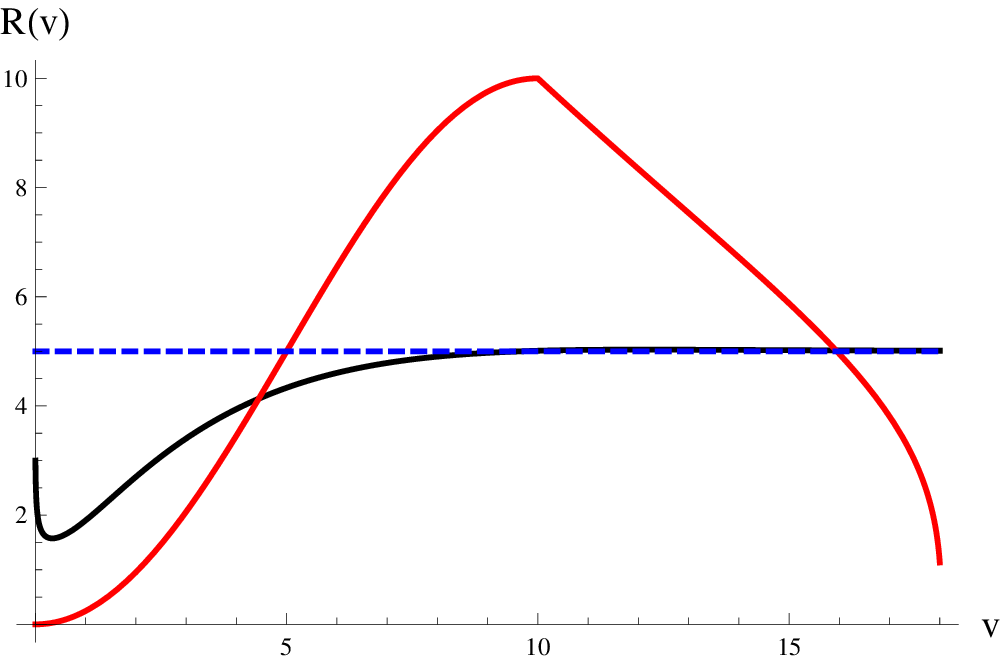}
\includegraphics[scale=0.7]{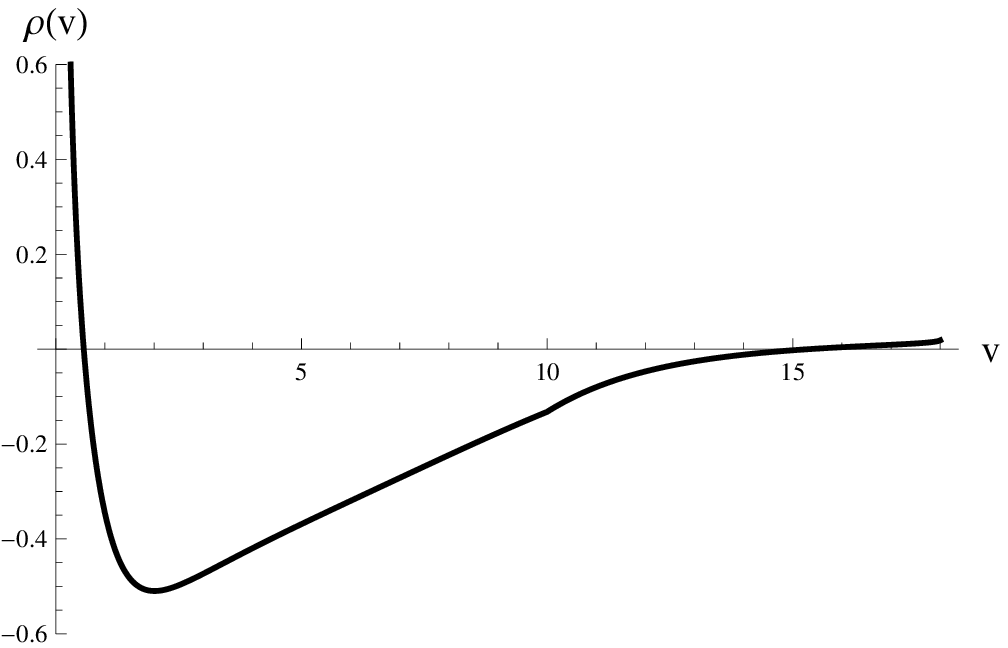}
\includegraphics[scale=0.7]{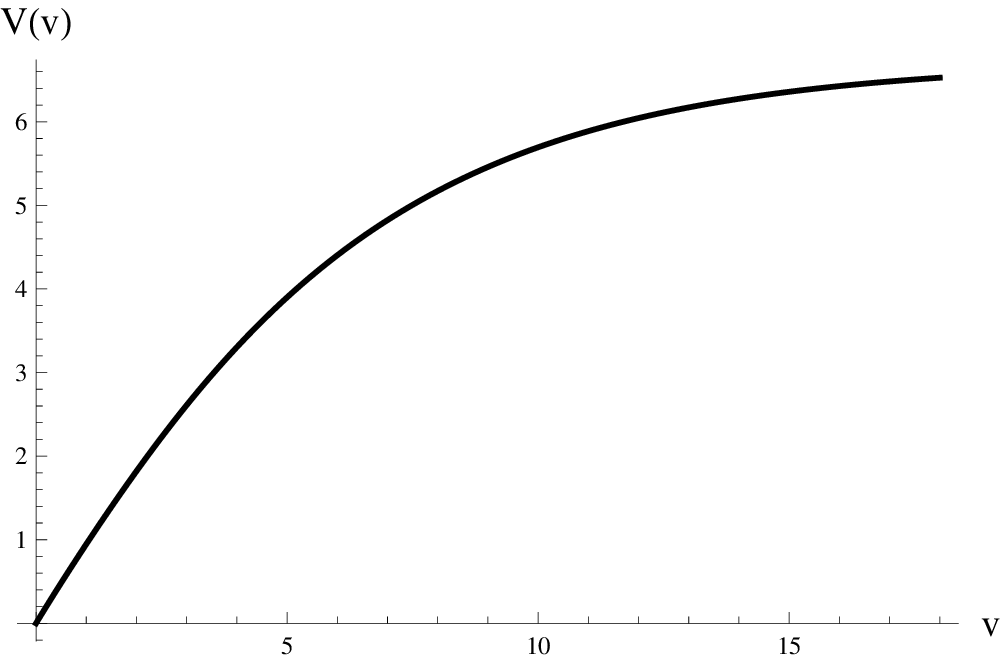}
\includegraphics[scale=0.7]{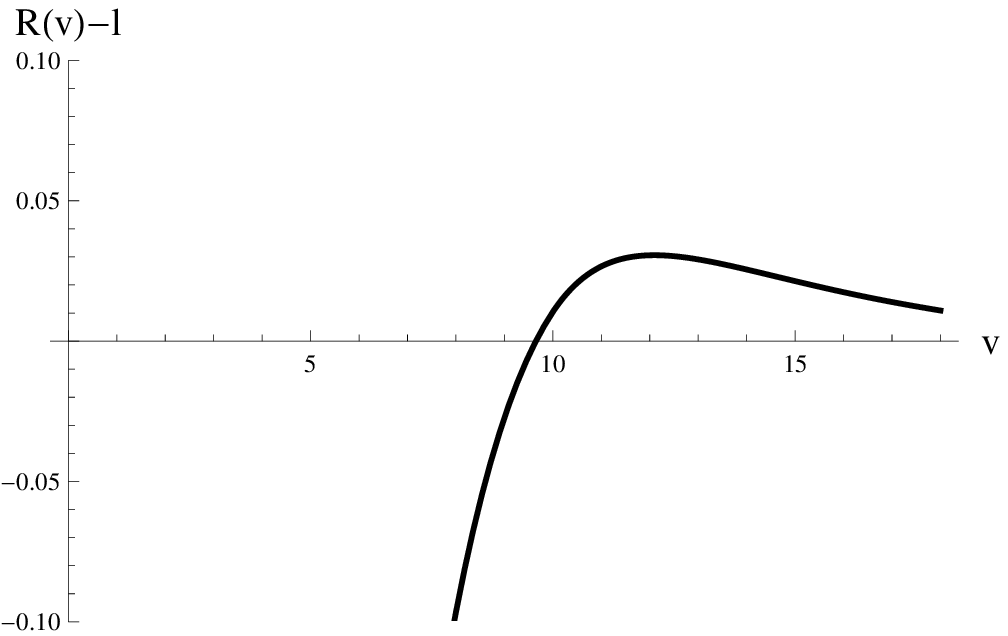}
\caption{\label{fig:type1_2}Numerical solution for $m_{0} = 5$, $\ell = 5$, $v_{0}=10$, $\alpha = 50$, $R(0) = 3$, $R'(0) = -100$, and $V(0) = 0$. Due to the choice if $R'$, the shell can cross $r = \ell$ surface. In this case, there appear outer and inner apparent horizons.}
\end{center}
\end{figure}

Figs.~\ref{fig:type1}, \ref{fig:type1_2}, and \ref{fig:type2} illustrate typical numerical results. Qualitatively, one can categorize that Figs.~\ref{fig:type1} and ~\ref{fig:type1_2} correspond to the strong Hawking radiation case (the large $\alpha$ limit) while Fig.~\ref{fig:type2} corresponds to the weak Hawking radiation case. On the other hand, Figs.~\ref{fig:type1} and \ref{fig:type2} correspond to the initially stationary shell $R'(0) = 0$ while Fig.~\ref{fig:type1_2} corresponds to the dynamical case ($R'(0) < 0$), where all other initial conditions are the same as those of Fig.~\ref{fig:type1}.

\subsection{Analysis: before evaporation}
\begin{figure}
\begin{center}
\includegraphics[scale=0.7]{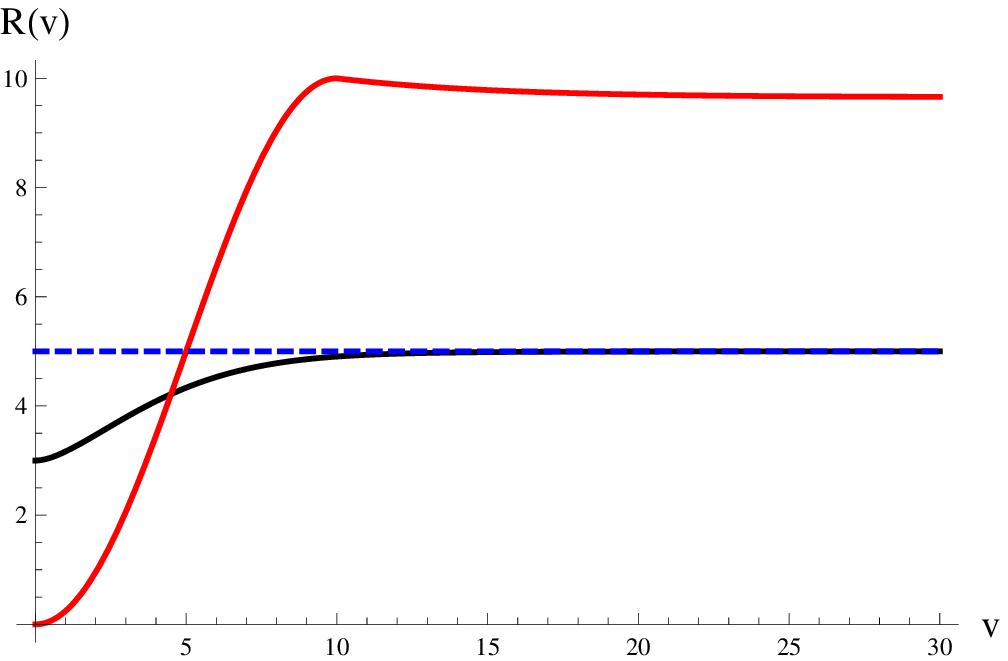}
\includegraphics[scale=0.7]{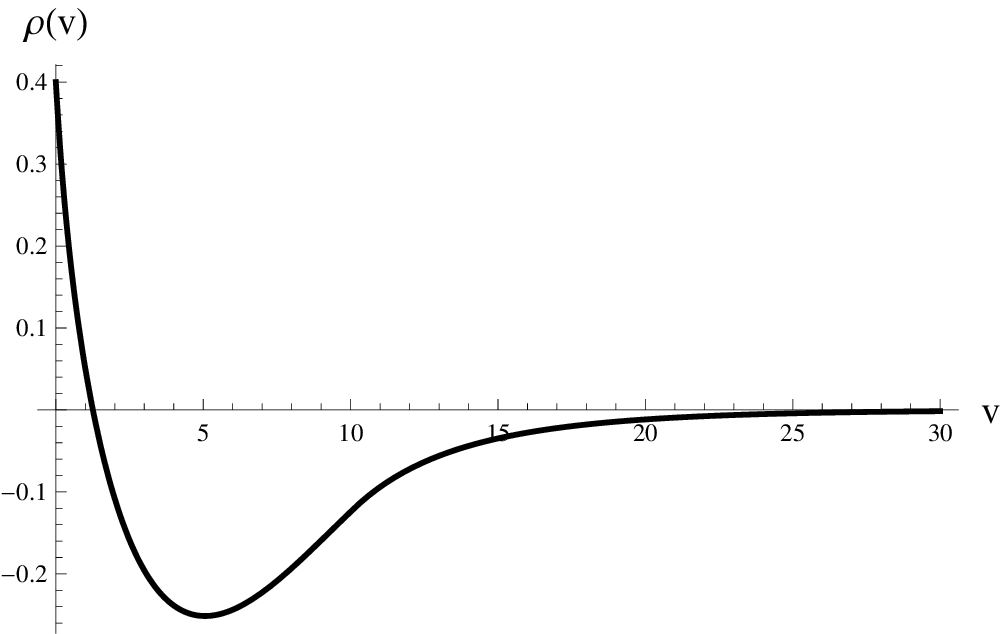}
\includegraphics[scale=0.7]{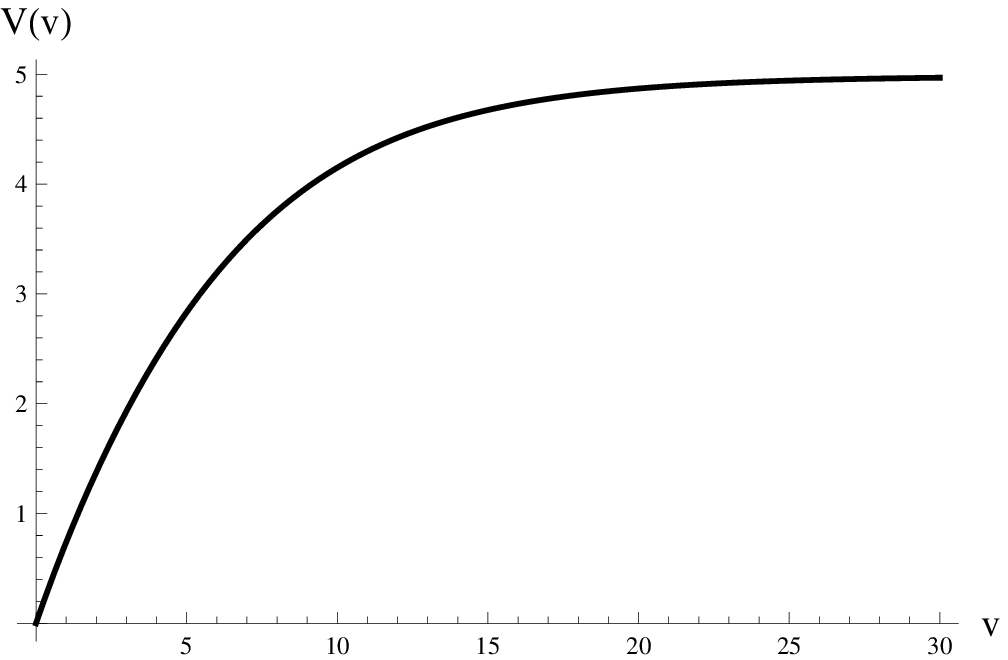}
\includegraphics[scale=0.7]{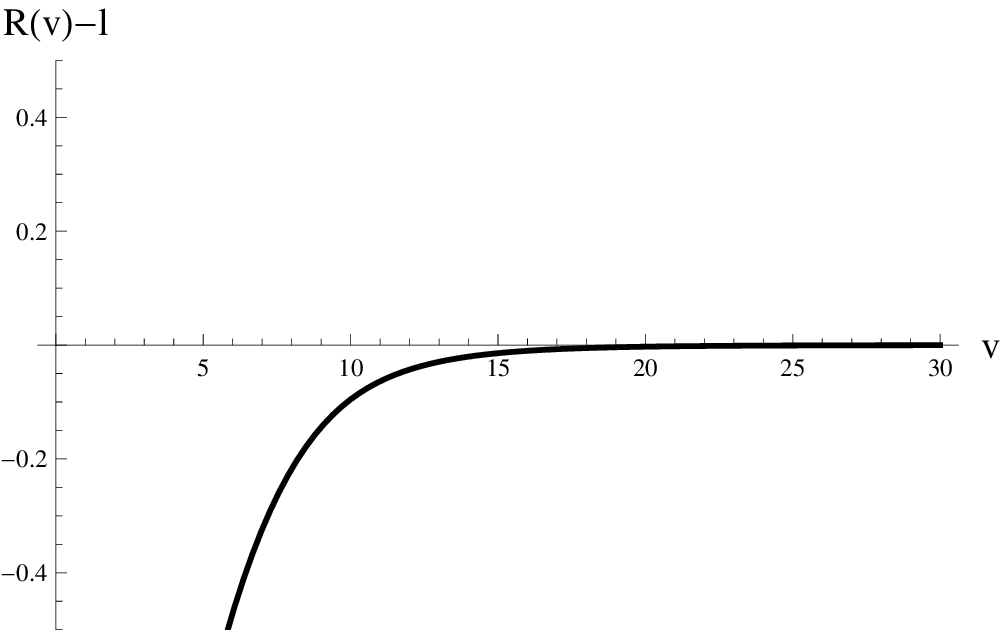}
\caption{\label{fig:type2}Numerical solution for $m_{0} = 5$, $\ell = 5$, $v_{0}=10$, $\alpha = 5$, $R(0) = 3$, $R'(0) = 0$, and $V(0) = 0$. The other results are similar to the previous result, but now the apparent horizon is way outside the shell.}
\includegraphics[scale=0.7]{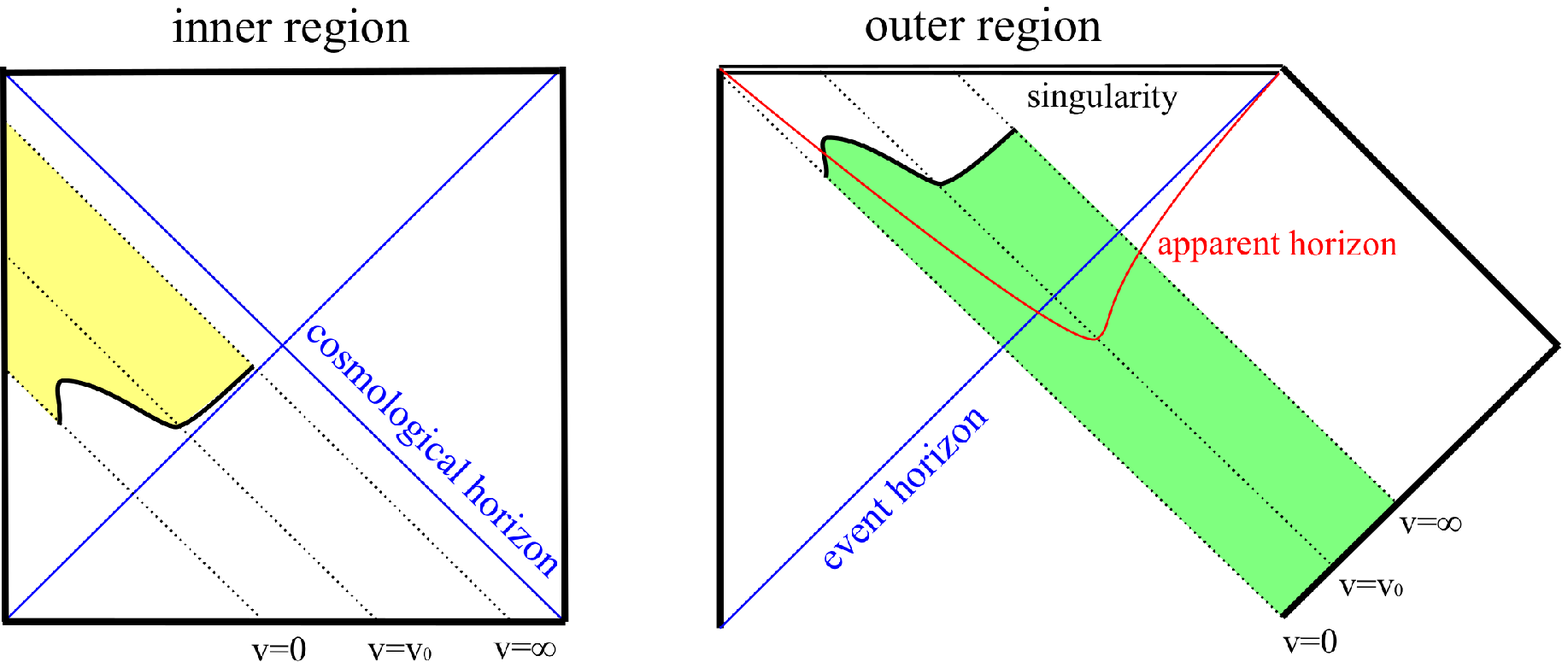}
\caption{\label{fig:causal}Causal structure for the most natural initial conditions, where inside (left) is de Sitter space, outside (right) is Vaidya metric, and the thick black curve is the location of the shell. Only the yellow colored region for left and the green colored region for right are physical.}
\end{center}
\end{figure}

We point out several interesting points during the collapsing phase ($v \leq v_{0}$):
\begin{itemize}
\item[--] At the beginning of collapse, the apparent horizon is completely inside the thin shell. Consequently, for the matter collapse, there is no apparent horizon up to a certain critical mass. This kind of initial condition is a typical assumption of the regular black hole picture \cite{Frolov:1988vj,Hayward:2005gi}; however, in the existing literature, authors did not assume a physical false vacuum bubble with collapsing matter. Rather, they extrapolated a limit of this scenario to consider a regular black hole solution with a de Sitter-like core. In this sense, our setup is more physical compared to previous literature.
\item[--] Above the critical mass, the apparent horizon grows outside the shell.
\item[--] In the beginning of this dynamical phase, the thin shell is time-like. In order to measure the signature of the shell, one can assume $d\Omega = 0$ and divide by $dVdv$ both sides of Eq.~(\ref{eq:metout}). Since the sign of the metric is related to the signature of the shell, one can measure this by calculating the quantity
\begin{eqnarray}
\rho \equiv - R' + \frac{1}{2} \left( 1 - \frac{2m}{R} \right) V'\,,
\end{eqnarray}
where if $\rho > 0$, the shell is time-like and if $\rho < 0$, the shell is space-like. During the gravitational collapse, the shell becomes \textit{space-like} and approaches $r = \ell$.
\item[--] By comparing Figs.~\ref{fig:type1}, \ref{fig:type1_2}, and \ref{fig:type2}, one can see that these behaviors do not sensitively depend on the choice of the initial conditions.
\end{itemize}

We note an interesting observation that the shell can become space-like during the gravitational collapse. Such a space-like, or \textit{tachyonic}, distribution of the shell can be interpreted an acausal transition of matter which is indeed very harmful if information is attached to the shell \cite{Chen:2017pkl}. On the other hand, if it is just interpreted as a gradient of some field, then it is not at all surprising that there appears a such a space-like distribution of the field and one can still do a sensible analyses of the dynamics \cite{Balbinot:1990zz}. Indeed, such a space-like distribution was originally expected albeit this fact was not much emphasized in the literature. Recently, effective black-to-white hole models have been severely constrained by considering effects of the thin-shell turning space-like \cite{BenAchour:2020bdt}. If we start from a static solution, the modified effective matter inside the horizon must be space-like. As one naturally extend the solution to the dynamical case, it is necessary that the intermediate matter turns into  space-like distribution at some instant \cite{Yeom:2008qw}. This is the novelty of our work which has been not been rigorously investigated up to now.

Finally, we summarize the causal structure of both inside and outside the shell in Fig.~\ref{fig:causal}.

\begin{figure}
\begin{center}
\includegraphics[scale=0.7]{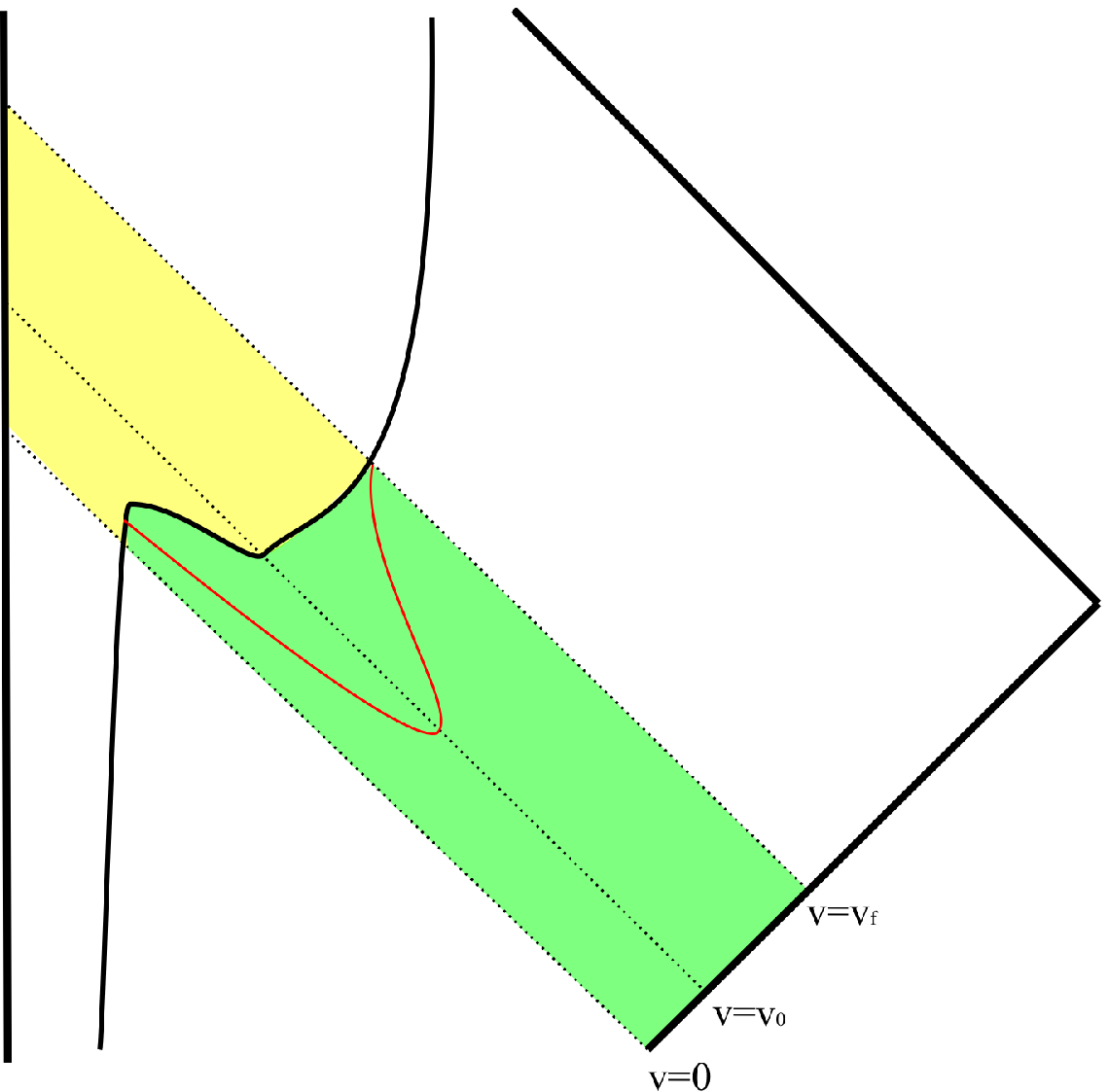}
\caption{\label{fig:causal4}The causal structure of the rapidly evaporating black hole.}
\end{center}
\end{figure}

\begin{figure}
\begin{center}
\includegraphics[scale=0.7]{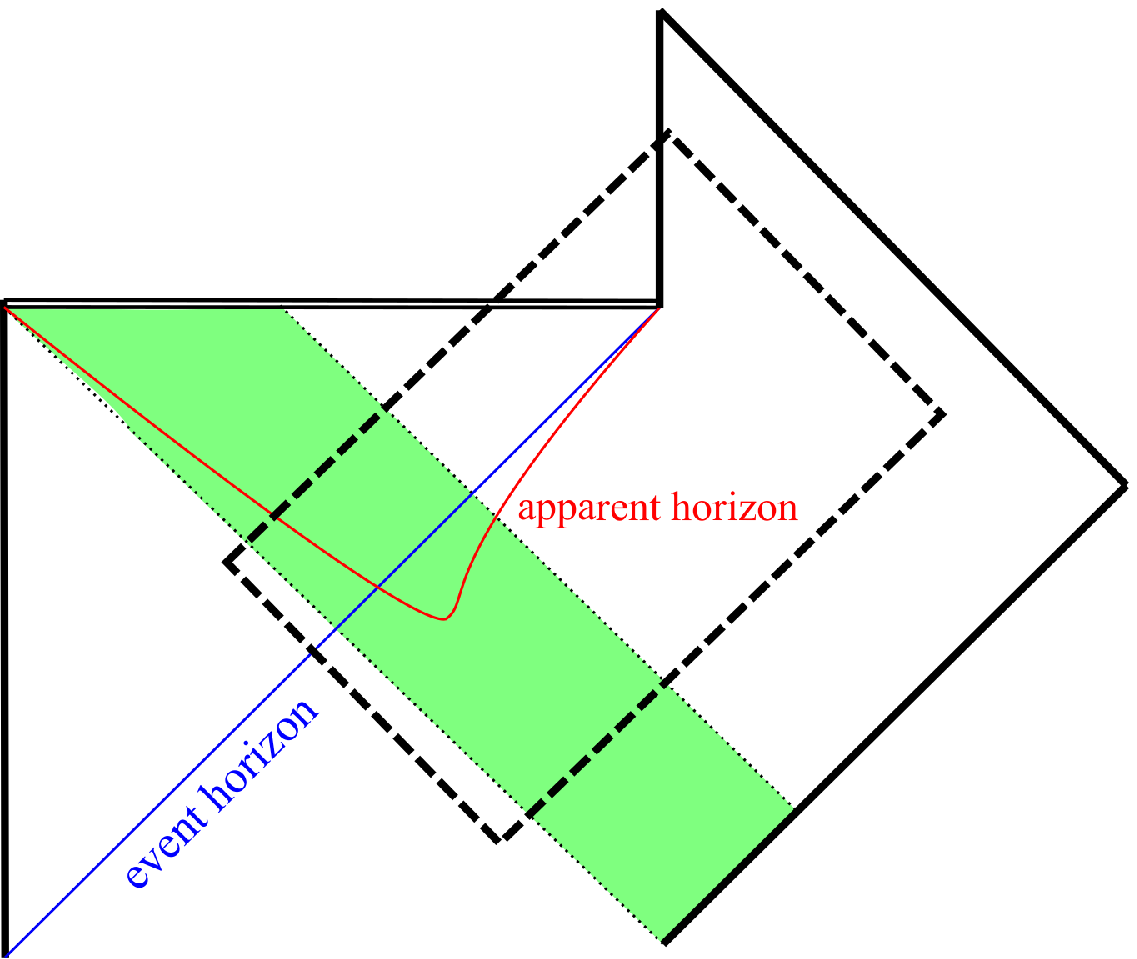}
\caption{\label{fig:causal3}Causal structure of an evaporating Schwarzschild black hole.}
\includegraphics[scale=0.7]{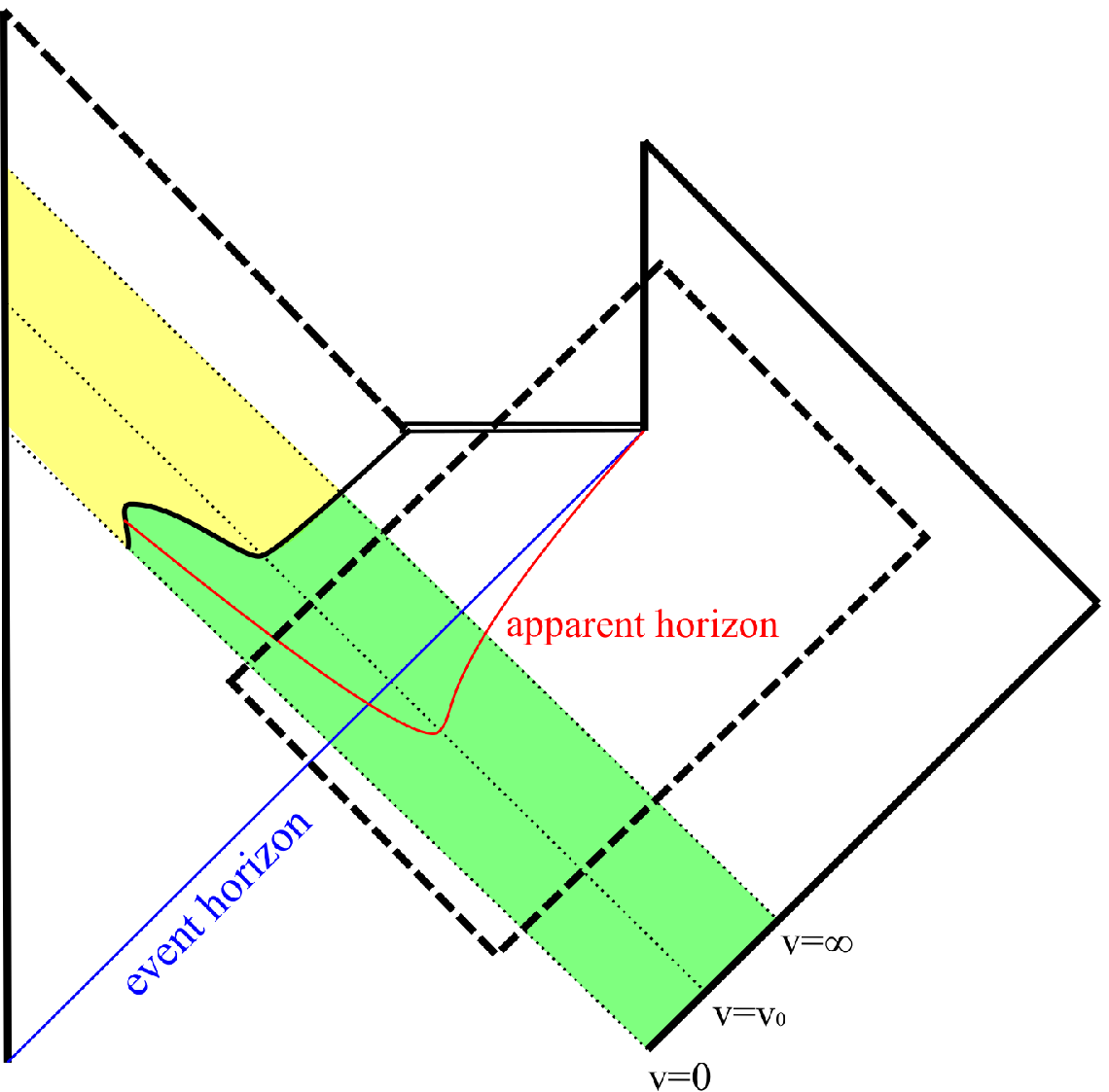}
\caption{\label{fig:causal2}The final result for the entire spacetime. The black dashed box corresponds to the causal past of the thin-shell which gives the same causal structure of Fig.~\ref{fig:causal3}.}
\end{center}
\end{figure}

\subsection{Analysis: during evaporation}
At the end of the gravitational collapse, there are two types of solutions:
\begin{itemize}
\item[Type 1:] If $\alpha$ is very large (Figs.~\ref{fig:type1} and \ref{fig:type1_2}), then the apparent horizon can shrink quickly so that the black hole disappears. After the black hole loses its mass below the critical mass, there remains an outgoing time-like shell (Fig.~\ref{fig:causal4}). This is qualitatively consistent with \cite{Hayward:2005gi} and \cite{Hwang:2012nn}.
\item[Type 2:] If $\alpha$ is small (Fig.~\ref{fig:type2}), then the shell approaches the null direction and within a finite coordinate time $V$, the parameter $v$ reaches to infinity as the shell approaches to the cosmological horizon of the internal de Sitter space. Hence, we need to choose a new coordinate time for the extension.
\end{itemize}

Even though our coordinate system breaks down, there is a reasonable way to extend the geometry. Since (1) the spacetime outside the shell is the same as Schwarzschild and (2) the shell approaches the null direction, the causal past of the shell will be determined by the usual black hole spacetime (black dashed box of Fig.~\ref{fig:causal3}. By consistently pasting Fig.~\ref{fig:causal} and Fig.~\ref{fig:causal3}, we obtain the entire spacetime of dynamical regular black holes in the weak Hawking radiation limit (Fig.~\ref{fig:causal2}). This is in keeping with expectations from numerical calculations \cite{Hwang:2012nn}.

Finally, we come to the case of the `rapid' evaporation of black holes. The lifetime of a black hole is of the order of $t \sim M^{3}/\alpha$, where $M$ is the mass of the black hole and $\alpha$ is a constant which is proportional to the number of matter fields that contribute to Hawking radiation. If $\alpha$ is large enough, a black hole can evaporate even before the formation of the singularity since the time scale for the latter is approximately $\tau \sim M$. If two time scales $t$ and $\tau$ are the same order, i.e., $t \sim \tau$, or equivalently, $\alpha \sim M^{2}$, then one can say that the black hole evaporates before a singularity is even formed. Figs.~\ref{fig:type1} and \ref{fig:type1_2} illustrate such cases, although this limit is not semiclassical in typical scenarios \cite{Yeom:2009zp}. Therefore, one may alternatively say, rather conservatively, that semiclassical regular black holes with a dS core should have a singularity as shown in the causal structure Fig.~\ref{fig:causal3}.

\section{\label{sec:qgc}Quantum geometry corrections}
It is natural to speculate how our analysis would change if one includes quantum gravity corrections beyond assuming a regular core as was done above. In order to discuss some of the typical corrections which arise in the effective spacetime of regular black holes from the underlying quantum geometry, let us focus on the specific case of loop quantum gravity. In this case, nonperturbative corrections from the theory arise in the form of having bounded (extrinsic) curvature due to regularization of the phase space in terms of SU$(2)$ holonomy matrix-elements. The typical behavior of black holes in loops is that one transitions from a black to white hole geometry, at a transition surface deep inside the core of the black hole. The field variables in this case do not satisfy Einstein's equations but are rather subject to effective equations which include several nonperturbative corrections. The details of this process are unimportant for our purposes and we shall only focus on a few qualitative features of such regular black holes.

Solving the vacuum effective equations in loop quantum gravity gives rise to regular static solutions which go beyond GR. However, in order to compare the resulting solutions with their classical counterparts, a common practice is to rewrite the quantum geometry corrections to the Einstein's equations in terms of an `effective' stress energy tensor. In this way, the non-vanishing of the Einstein tensor, construed as this effective stress energy tensor (perhaps, of some anisotropic fluid), can then be compared with the vacuum solution of GR which in this case is given by Schwarzschild metric. The typical LQG solutions \cite{Bojowald:2018xxu} lead to the violation of the strong energy condition \cite{Bouhmadi-Lopez:2020oia} in the deep quantum regime. Near the transition surface, this violation leads to large departures from GR and makes the curvature invariants approach an upper maximum value. However, this quantum geometry stress energy component can be seen to die down quickly as one approaches the horizon and becomes negligibly small outside it. Thus, singularity resolution in these models leads to an effective metric such that the strong energy condition is violated in a large neighborhood of the transition surface from black to white holes, even for static solutions.

Although, thus far, only static black hole solutions have been studied in loop quantum gravity, it would be interesting to speculate how such nonperturbative corrections would affect the dynamical situation. As we have seen in this work, in order to have singularity resolution in a dynamical regular black hole, one requires (perhaps, unnaturally) rapid evaporation as can be seen in Fig.~\ref{fig:causal2}. Since Hawking radiation implies a negative energy contribution, in effect one requires \textit{sufficiently} violating the energy conditions in order to evade singularities. This is precisely where the loop quantum gravity effects might come in handy. For such regular black holes, the dynamical evaporation might not need to be similarly strong due to addition negative energy flux due to quantum geometry effects coming from within the cores. The main idea would be that nonperturbative quantum effects would ameliorate the need for very rapid evaporation of regular black holes in order to have complete singularity resolution. These types of solutions would be studied in future within the thin shell formalism.

\section{\label{sec:con}Conclusion}
In this paper, we investigated a regular black hole model which has a dS-like core. In regular black holes, there should be a transient region that connects from the outside black hole to the regular center. In order to describe the transient region consistently, we need a model that can include field theoretical dynamics. (Such a field theoretical construction has largely been ignored in the literature thus far.) If the center is akin to dS space, then, in principle, one can mimic the model with such a field theoretical structure. In this work, we have used the thin-shell approximation and considered dynamics of the shell by using the Vaidya metric. Due to this technical improvement, we can regard the transient region dynamically with justifiable methods.

This results in revealing several interesting aspects. During the gravitational collapse phase, the shell can move along the space-like direction. During the black hole evaporation phase, the shell and the outer apparent horizon will approach each other, but unless the evaporation is very rapid, the approaching tendency is too slow. Therefore, it is reasonable to conclude that although a false vacuum core may remove singularity along the incoming null direction, there still exists a singularity along the outgoing null direction, unless the evaporation is very strong. Therefore, we can conclude that the effective modification of the core region is not enough to explain the resolution of the singularity in a genuinely dynamical scenario. We need something more to explain singularity-resolution in the out-going null direction within the semiclassical approximation, when the evaporation rate is reasonably slow. The closed apparent horizon is indeed relied on these subtle assumptions that need more justifications.

In future work, there may be two ways to generalize this model:
\begin{enumerate}
\item \textit{Non-zero tangential stress}: Note that the physical parameters (e.g., tension and tangential stress) of the shell should come from the detailed origin of the field and the underlining potential structure that form the shell. However, in our formalism, the tension is geometrically determined via Einstein's equations. The important assumption is the zero tangential stress condition; if we do not insist on this condition any longer, then there can be other possibilities. Therefore, it is reasonable to revisit the vanishing tangential stress condition. One straightforward possibility to generalize this would be to choose
\begin{eqnarray}
{G^{\theta}_{\;\theta}}_{\mathrm{shell}} = w {G^{v}_{\;v}}_{\mathrm{shell}}\,,
\end{eqnarray}
where $w$ is the equation of state.
\item \textit{Internal structure}: There may be some variations of the internal geometry rather than assuming a pure de Sitter core, e.g., adopting the Hayward model or further such modifications. Although unlikely, this may help in regularizing the mass inflation instability.
\end{enumerate}
Eventually, if one can build a regular black hole model which can justify its dynamics based on field theoretical calculations, it shall be successful in resolving the singularity problem, at least within the semiclassical approximation. On the other hand, if such an endeavor is proven to be impossible, then we need to take an alternate route to explain the singularity issue, for instance, by introducing the wave function and going beyond effective spacetimes. We leave all of these interesting issues for future investigations.

\section*{Acknowledgment}
The authors would like to thank helpful discussions with Gabor Kunstatter, William Unruh, and Pisin Chen. This work was supported by the Korean Ministry of Education, Science and Technology, Gyeongsangbuk-do and Pohang City for Independent Junior Research Groups at the Asia Pacific Center for Theoretical Physics and the National Research Foundation of Korea (Grant No.: 2018R1D1A1B07049126). SB is also supported in part by funds from NSERC, from the Canada Research Chair program, by a McGill Space Institute fellowship and by a generous gift from John Greig.

\end{document}